%% file: arxiv.tex
\documentclass[sigconf,nonacm]{acmart}

\setcopyright{none}
\renewcommand\footnotetextcopyrightpermission[1]{}

\usepackage{amsmath}
\usepackage{booktabs}
\usepackage{graphicx}
\usepackage{microtype}
\usepackage{dblfloatfix}

\input{tables/generated_numbers}

\newcommand{\HNNMarg}{\textsc{HNN (marginal)}}
\newcommand{\HNNMS}{\textsc{HNN (m-s)}}
\newcommand{\HNNShuffle}{\textsc{HNN (input-shuffled)}}
\newcommand{\MLPHNN}{\textsc{MLP-HNN}}
\newcommand{\NNThree}{\textsc{NN3}}

\begin{document}

\title{Dependence-Informed Sparse Neural Architecture for Stock Return Prediction}

\author{Hongyu Lin}
\affiliation{%
  \institution{University College London}
  \city{London}
  \country{United Kingdom}
}

\author{Yulin Chen}
\affiliation{%
  \institution{University College London}
  \city{London}
  \country{United Kingdom}
}

\author{Yuanrong Wang}
\affiliation{%
  \institution{University College London}
  \city{London}
  \country{United Kingdom}
}

\author{Antonio Briola}
\affiliation{%
  \institution{University College London}
  \city{London}
  \country{United Kingdom}
}

\author{Tomaso Aste}
\affiliation{%
  \institution{University College London}
  \city{London}
  \country{United Kingdom}
}

\begin{abstract}

Using neural networks for stock return prediction typically requires choices about depth and hidden-layer width that are difficult to connect to financial interpretation. We study an alternative: estimate dependence among firm characteristics with a Maximally Filtered Clique Forest (MFCF), then map its clique structure to a Homological Neural Network (HNN). The MFCF maximum clique size
\(K\) is the only parameter controlling architectural complexity, and it has a clear graphical meaning: it bounds the number of characteristics in each maximal clique and hence the highest interaction order the network can represent. The filtered graph then fixes the neural network's depth, layer widths, and sparse connections before
training, in place of a separately chosen depth and width sequence. We apply two HNN variants to annual out-of-sample forecasts of U.S. stock excess returns from 1987 to 2016 using 94 firm characteristics. The HNN models match a three-hidden-layer benchmark on pooled predictive accuracy, rank the cross-section more accurately, and use roughly \ParamRatio{} times fewer parameters than a fully connected network with the same induced layer widths. Two structural ablations indicate that both the sparse connectivity and the
estimated grouping of characteristics contribute to the ranking advantage, and both effects remain significant after correcting for multiple testing. These findings show that HNNs offer a practical and interpretable way to incorporate estimated dependence among firm characteristics into neural architecture design.

\end{abstract}


\maketitle

\section{Introduction}

Machine learning allows empirical asset-pricing models to capture a broad range of nonlinear relationships. Gu, Kelly, and Xiu (GKX)~\cite{gu2020} show that neural networks and tree-based models outperform linear alternatives in forecasting cross-sectional returns, suggesting that higher-order interactions among firm characteristics contain useful predictive information. Related studies develop characteristic-based latent factor models~\cite{kelly2019}, nonlinear autoencoder and deep-learning asset-pricing models~\cite{gu2021autoencoder,chen2024}, regularized approaches to the cross-section~\cite{kozak2020}, nonparametric characteristic selection~\cite{freyberger2020}, and tests for identifying incremental pricing
factors~\cite{feng2020}.
Yet using a multilayer perceptron (MLP) still requires researchers to specify
its depth and hidden-layer widths. These choices are generally selected through
validation and have limited direct financial interpretation~\cite{elsken2019}.

We investigate whether a neural architecture can instead be constructed from estimated dependence among firm characteristics. Characteristics related to valuation, past returns, liquidity, investment, and profitability exhibit substantial empirical dependence~\cite{kozak2020,feng2020}. To obtain a sparse representation of these relationships, we apply Information Filtering Networks (IFNs) to the training data. IFNs filter a dense dependence matrix under topological constraints~\cite{mantegna1999,tumminello2005,massara2017}. Specifically, the Maximally Filtered Clique Forest (MFCF) organizes the retained relationships into cliques of bounded size while preserving graph decomposability~\cite{massara2019}. A Homological Neural Network (HNN) maps these cliques to a neural architecture: characteristics form the input layer, retained pairs form the second-order layer, larger subsets form subsequent layers, and connections follow set inclusion~\cite{wang2023hnn,lin2026}. The number of layers is therefore the highest interaction order the filtered graph attains, rather than a capacity setting chosen in advance. A conventional MLP uses fully connected layers and may apply \(l_1\) regularization to reduce the influence of unhelpful connections~\cite{tibshirani1996,gu2020}. In contrast, the HNN is sparse by construction: MFCF determines which connections are included before the network is trained. This architecture follows the compositional view that high-dimensional functions can be assembled from lower-dimensional constituents~\cite{poggio2017}.

We evaluate this approach using the GKX sample design~\cite{gu2020} and 94
ranked firm characteristics to predict subsequent monthly excess returns. We
consider two HNN variants: \HNNMarg{}, which constructs the network using
dependence among characteristics alone, and \HNNMS{}, which also uses training
returns to distinguish between above- and below-median observations. Both are compared with \NNThree{}~\cite{gu2020}, a compact three-hidden-layer
network that performs best among the models GKX evaluate, adapted
to the same information set.
 For ablation studies, an input-shuffling HNN tests
whether the alignment of characteristics with the MFCF-derived graph matters,
while a fully connected network with the same induced layer widths provides a
dense architectural comparison. Evaluation covers both level prediction and cross-sectional ranking. We measure
ranking with information coefficients (ICs) and returns from prediction-sorted
portfolios. The Pearson and Spearman ICs are monthly cross-sectional
correlations between predicted and realized returns. Spearman IC measures the ordering of stocks, whereas Pearson IC also reflects
the linear association between the magnitudes of predicted and realized
returns. Prediction-sorted portfolios test whether this cross-sectional
ordering produces economically meaningful return separation.

The study makes three empirical contributions. First, both HNN variants achieve
numerically higher pooled out-of-sample \(R^2\) than \NNThree{}, and \HNNMarg{}
attains the highest Pearson IC and equal-weighted portfolio spread of any model
we consider, together with the highest Spearman IC among the neural models.
Second, against a fully connected network matched on hidden-layer widths, the
HNN reduces the parameter count by a factor of \ParamRatio{} while improving
both information coefficients and the equal-weighted spread. Third, randomly
permuting the input characteristics with the architecture and capacity held
fixed weakens cross-sectional ranking, so the alignment between characteristics
and their inferred clique structure contributes beyond sparsity alone. After Holm
adjustment for multiple comparisons~\cite{holm1979}, the Pearson IC differences
remain significant, while on \(R^2\), Spearman IC, and portfolio spread \HNNMarg{} and \NNThree{} are statistically indistinguishable.


The remainder of the paper is organized as follows. Section~\ref{rela} reviews
machine learning in asset pricing and prior work on filtered networks and
structured neural models. Section~\ref{method} develops the HNN construction,
and Section~\ref{exp} describes the data, walk-forward protocol, benchmarks, and
evaluation procedure. Section~\ref{re} reports predictive accuracy,
cross-sectional ranking, parameter efficiency, transaction-cost sensitivity, and
stability over time. Sections~\ref{dis} and~\ref{con} discuss limitations and
conclude.

\section{Related Work}\label{rela}

\paragraph{Machine learning in asset pricing.}
GKX~\cite{gu2020} compare a wide range of methods on a large U.S. equity panel
and find that flexible nonlinear models improve out-of-sample return prediction,
partly by capturing interactions among predictors. Freyberger et
al.~\cite{freyberger2020} use regularization for nonparametric characteristic
selection, and Kozak et al.~\cite{kozak2020} combine shrinkage with
principal-component structure. Related latent-factor models use characteristics
to parameterize time-varying factor exposures~\cite{kelly2019}; conditional
autoencoders make those exposures nonlinear in the
characteristics~\cite{gu2021autoencoder}; and Chen et al.~\cite{chen2024} fit
nonlinear models subject to no-arbitrage restrictions. Where these papers use a
neural network, its depth and widths are chosen by hand.

The flexibility of these methods increases the importance of regularization,
model selection, and careful out-of-sample evaluation, particularly when  signals are weak and differences across models are modest.
Work on factor selection and multiple testing further motivates disciplined
comparisons~\cite{feng2020,harvey2016}. Accordingly, we use a common information
set and walk-forward protocol across models, together with multiplicity-adjusted
inference~\cite{holm1979} for the main architecture comparisons.

\paragraph{Filtered networks and Homological Neural Networks.}
Information Filtering Networks extract a sparse network of the strongest
relationships from a dense dependence
matrix~\cite{tumminello2005,massara2017}. Sparsity comes from a global criterion
applied under a constraint on the graph topology, rather than from penalizing
edges individually. The choice of constraint decides how much higher-order
structure survives: a spanning tree contains no clique larger than a
pair~\cite{mantegna1999}, while planar and chordal filters retain triangles and
larger cliques. The Maximally Filtered Clique Forest~\cite{massara2019} is of the latter kind and returns a forest of overlapping cliques of bounded size.

Filtered networks have been used before to study and forecast financial
dependence. Wang and Aste~\cite{wang2022icaif}, for example, build a filtered
graph and use it as the adjacency structure of a spatial-temporal graph neural
network, so the graph is an input to a model chosen separately. In the
Homological Neural Network (HNN) the filtered structure becomes the model
itself, with the cliques and their subsets as neural units~\cite{wang2023hnn}; later
work frames this as compositional sparsity and shows that, with the maximum
clique size held fixed, it matches or outperforms fully connected networks on
tabular regression with far fewer connections~\cite{lin2026}. This evidence comes
from problems with far more signal than the return cross-section, and whether a
structure estimated from the characteristics alone carries predictive
information here is untested. We apply the construction to a financial panel,
select the clique size by validation in each window so that depth follows from
the data, and assess the forecasts by portfolio returns as well as squared error.

\section{Method}
\label{method}

\subsection{Prediction Target and Characteristic Dependence}

Let \(x_{i,t}\in\mathbb{R}^{p}\) collect the \(p\) firm characteristics of
stock \(i\) at month \(t\). The prediction target is the stock's excess return
over the following month,
\begin{equation}
 y^{e}_{i,t+1}
 =
 R_{i,t+1}-R^{f}_{t+1},
 \qquad
 \widehat y^{e}_{i,t+1}
 =
 f_{\theta,\mathcal A}(x_{i,t}),
 \label{eq:prediction}
\end{equation}
where \(R_{i,t+1}\) is the realized return on stock \(i\) in month \(t+1\),
\(R^{f}_{t+1}\) is the one-month risk-free rate, \(\theta\) contains the
trainable parameters, and \(\mathcal A\) denotes the network architecture.

The architecture is estimated from dependence among firm characteristics. Let
\(\mathbf X\in\mathbb{R}^{N\times p}\) contain the \(N\) observations of the current training window, and let \(X_a\) denote its \(a\)-th column,
one per characteristic. The characteristics are the nodes of the graph,
\(\mathcal V=\{1,\ldots,p\}\), and the resulting network is fixed across all firms and dates. The filter takes as input the absolute Pearson correlation matrix 
\begin{equation}
 D_{ab}
 =
 \left|
 \operatorname{corr}(X_a,X_b)
 \right|,
 \qquad a,b\in\mathcal V .
 \label{eq:dependence}
\end{equation}

Linear correlation is used rather than a more general dependence measure.
Because the characteristics enter as cross-sectional ranks,
Equation~\eqref{eq:dependence} is a correlation between rank variables and is
therefore insensitive to outliers and to monotone transformation of the
underlying characteristic~\cite{kendall1970}. It also avoids the extra tuning that measures such
as mutual information~\cite{cover2006elements} require, and richer measures can be substituted without
changing the rest of the construction~\cite{lin2026}.

The MFCF filters \(D\) into overlapping characteristic cliques while preserving
a decomposable graph structure~\cite{massara2019}. At each step, a
characteristic \(v\in\mathcal V\) not yet in the graph is attached through an
admissible separator \(S\), meaning a clique of the current graph whose
extension preserves decomposability, forming the new clique \(S\cup\{v\}\). The
attachment is scored by
\begin{equation}
 G(v,S)
 =
 \sum_{u\in S}D_{vu}^{2}.
 \label{eq:gain}
\end{equation}
Repeatedly selecting the largest admissible gain produces a collection of
maximal cliques \(\mathcal C\). Unlike sparse precision matrix estimators such as the graphical lasso~\cite{friedman2008}, which identify pairwise conditional dependence edges, the MFCF yields a filtered graph whose maximal cliques define overlapping groups of characteristics. In our implementation, the only structural parameter is the maximum clique size K, which imposes
\[
|C|\leq K
\qquad
\text{for every } C\in\mathcal C.
\]
The candidate values considered for K are reported in Section~\ref{exp}.

\begin{figure*}[t]
  \centering
  \includegraphics[width=\textwidth]
  {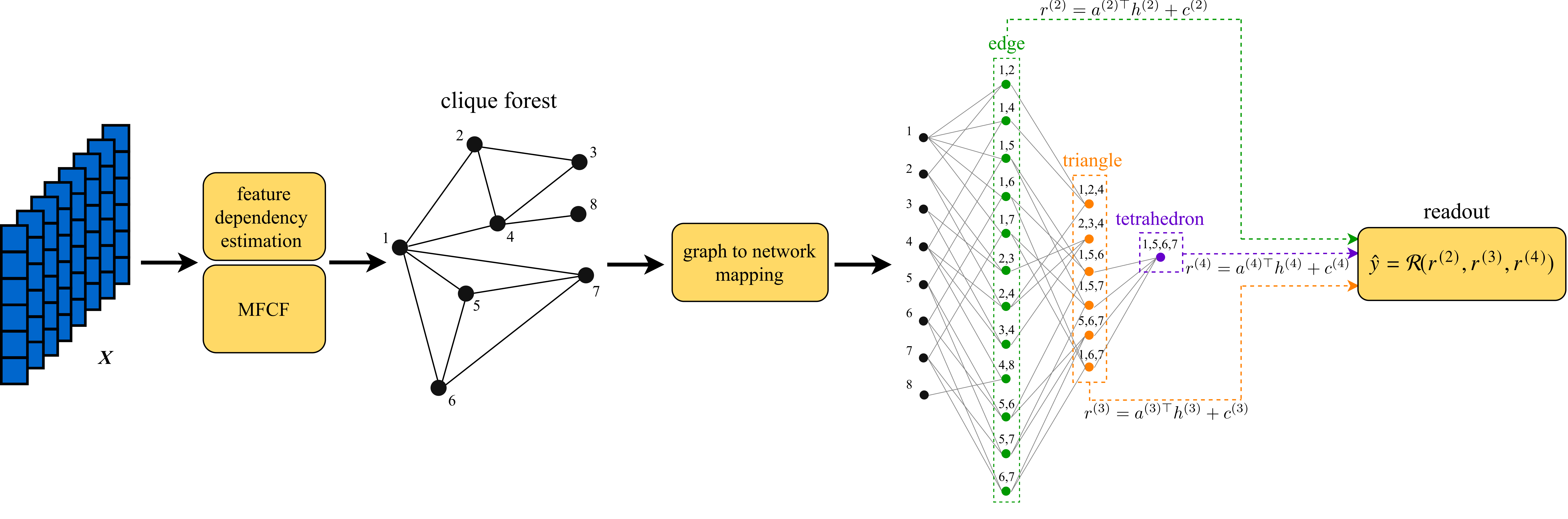}
  \caption{Dependence-informed HNN construction. MFCF filters training-only
  characteristic dependence into overlapping cliques. Retained subsets form
  ordered neural units, set inclusion determines sparse connections, and
  summaries from all non-input layers form the excess-return forecast. Adapted
  from Lin et al.~\cite{lin2026}.}
  \Description{A stock-month panel is converted into a characteristic
  dependence matrix and a clique forest, then into sparse neural layers
  representing edges, triangles, and a tetrahedron. All non-input layers
  contribute to the final prediction.}
  \label{fig:pipeline}
\end{figure*}

\subsection{Clique-Induced Neural Architecture}

Figure~\ref{fig:pipeline} illustrates how the MFCF is translated into an
HNN~\cite{wang2023hnn,lin2026}. The cliques and their subsets become the neural units of the network, and set inclusion becomes its connectivity.

Since \(K\) is an upper bound, the highest order the filtered graph actually
reaches is
\begin{equation}
 K^{\star}=\max_{C\in\mathcal C}|C|\leq K .
 \label{eq:kstar}
\end{equation}
Let \(\mathcal H^{(k)}\) collect the distinct order \(k\) interactions the filter
retains:
\begin{equation}
 \mathcal H^{(k)}
 =
 \left\{
 \alpha\subseteq\mathcal V:
 |\alpha|=k,\
 \alpha\subseteq C
 \text{ for some }C\in\mathcal C
 \right\}.
 \label{eq:complex}
\end{equation}
A subset that appears in several maximal cliques contributes a single unit.
Layer \(k\) has one unit for each \(\alpha\in\mathcal H^{(k)}\), so layer 1
holds the individual characteristics, layer 2 the retained pairs, and later
layers contain progressively larger cliques. Its width \(n_k=|\mathcal H^{(k)}|\) is
determined by the filtered graph rather than chosen, and \(K\) bounds the
number of layers, not their widths.

Connections follow inclusion: the unit for \(\alpha\in\mathcal H^{(k)}\)
receives input only from the units for its subsets of size \(k-1\).
Characteristics therefore interact only when the filter places them in a common
clique. With \(h^{(1)}_{i,t}=x_{i,t}\), the activations at order \(k\) are
\begin{equation}
 h^{(k)}_{i,t}
 =
 \operatorname{ReLU}\!\left(
 \operatorname{LN}_{k}\!\left[
 W^{(k)}h^{(k-1)}_{i,t}+b^{(k)}
 \right]\right),
 \quad k=2,\ldots,K^{\star},
 \label{eq:hnnlayer}
\end{equation}
where \(h^{(k)}_{i,t}\in\mathbb{R}^{n_k}\) holds the order-\(k\) activations,
\(b^{(k)}\in\mathbb{R}^{n_k}\) is a bias, \(\operatorname{LN}_{k}\) is layer
normalization~\cite{ba2016}, and ReLU is the rectified linear unit~\cite{nair2010}. The weight
matrix \(W^{(k)}\in\mathbb{R}^{n_k\times n_{k-1}}\) is sparse by construction:
the entry \(W^{(k)}_{\alpha\tau}\) is trainable when \(\tau\subset\alpha\) and
fixed at zero otherwise.

Conventional feed-forward networks predict from the last hidden layer. The HNN
instead reads out from every layer above the input. Each layer is first
summarized by a scalar,
\begin{equation}
 r^{(k)}_{i,t}
 =
 (a^{(k)})^\top h^{(k)}_{i,t}+c^{(k)},
 \qquad k=2,\ldots,K^{\star},
 \label{eq:layer-readout}
\end{equation}
and the summaries are combined to form the forecast,
\begin{equation}
 \widehat y^{e}_{i,t+1}
 =
 \sum_{k=2}^{K^{\star}}\beta_k r^{(k)}_{i,t}+d,
 \label{eq:hnn-readout}
\end{equation}
where \(a^{(k)}\in\mathbb{R}^{n_k}\) and the scalars \(c^{(k)}\), \(\beta_k\)
and \(d\) are trainable.
 Every order therefore has a direct path to the
forecast, which matters because a unit belonging to no larger clique would
otherwise never reach the output. The linear form also keeps the contribution
of each order separable. The MFCF fixes the units and the connections before
training, so training estimates only the nonzero weights.

Each unit at order \(k\) has exactly \(k\) incoming connections, one for each of
its subsets of size \(k-1\). The HNN therefore carries far fewer weights than a
fully connected network with the same layer widths:
\begin{equation}
 P_{\mathrm{sparse}}
 =
 \sum_{k=2}^{K^{\star}}k\,n_k,
 \qquad
 P_{\mathrm{dense}}
 =
 \sum_{k=2}^{K^{\star}}n_{k-1}n_k .
 \label{eq:parameter-growth}
\end{equation}
The sparse count grows linearly in the layer widths, whereas the dense count
grows with their products.

\subsection{HNN Variants and Structural Controls}
The two variants use the construction above and differ only in the dependence
matrix supplied to MFCF.
\paragraph{\HNNMarg.}
The marginal specification estimates a single dependence matrix from all
observations in the training window. Returns are not used, so the architecture
depends only on the characteristics.

\paragraph{\HNNMS.}
The median-split specification splits the training observations at the median excess return. A dependence matrix and an MFCF are estimated within each subsample,
giving clique collections \(\mathcal C^{+}\) and \(\mathcal C^{-}\), and the
network is built from their union,
\begin{equation}
 \mathcal C
 =
 \mathcal C^{+}\cup\mathcal C^{-}.
 \label{eq:ms-union}
\end{equation}
Equation~\eqref{eq:complex} then applies unchanged, so a subset occurring in
both collections still contributes a single unit. Validation and
test returns are never used.

The two controls start from \HNNMarg{}: the first breaks the alignment
between characteristics and the MFCF-derived architecture, the second replaces the sparse HNN with a width-matched dense architecture.

\paragraph{\HNNShuffle.}
This ablation permutes the characteristics across the inputs of the \HNNMarg{} network, with no characteristic left at its original place. The same
permutation is used for every observation in a training window. Since each
input corresponds to a node of the filtered graph, this shuffles the
characteristics across the nodes while leaving the graph itself untouched: the
topology, the neural units, the connections, and the parameter count stay exactly as in
\HNNMarg{}, but the characteristics are no longer aligned with their original nodes in the MFCF. Capacity is unchanged, so a difference in performance is attributable to
that mismatch alone.

\paragraph{\MLPHNN}
This control keeps \HNNMarg{}'s depth and layer widths but connects every layer
fully and predicts from the last layer instead of from all of them. Because its connectivity and prediction head follow a conventional
MLP design, it provides an architectural comparison rather than a strict
connectivity ablation. The widths are matched but the parameter counts are not,
so it is substantially larger than \HNNMarg{}.

Together, these four models answer three questions: whether using returns to
estimate the graph helps (\HNNMS{} against \HNNMarg{}), whether characteristics must remain aligned with the MFCF (\HNNShuffle{}), and how the HNN architecture compares with a conventional dense network having the same
induced depth and layer widths (\MLPHNN{}).

\section{Experimental Design}
\label{exp}

\subsection{Data and Walk-Forward Protocol}

We use the GKX U.S. equity panel of 94 firm characteristics and monthly stock
returns from 1957 to 2016~\cite{gu2020}. Returns are taken in excess of the
contemporaneous risk-free rate. Within each month the observed characteristics
are ranked across stocks and mapped to \([-1,1]\), and missing values are set to
zero, the neutral rank. A near-zero-variance screen applied to each window's
training data leaves between 90 and 94 characteristics, so the MFCF decides
how characteristics are connected rather than which are available. The correlation matrix is computed from 150{,}000 training observations drawn at
random in each window.
The test sample contains \NTest{} observations.

Following GKX~\cite{gu2020}, the training window starts in 1957 and expands
by one year at each step, the following twelve years serve as validation, and
the next year is the test year. The first test year is 1987 and the last 2016,
giving 30 non-overlapping test years. Preprocessing, graph construction,
hyperparameter selection and fitting are redone in each window on that window's
training and validation data alone.

We do not reproduce GKX's full specification: all models start from the same 94
characteristics, without industry dummies or macroeconomic interactions. Fixing
the information set means differences in performance can be read as differences
in model construction.

\subsection{Models and Selection}

Seven models are compared. The four HNN specifications are defined in
Section~\ref{method}; the other three are benchmarks from GKX~\cite{gu2020}.
Huber-3 is their OLS-3: a regression on size, book-to-market and 12-month
momentum, fitted under Huber loss~\cite{huber1964} to limit the influence of
extreme returns. It shows what a minimal, long-standing specification achieves
on this panel. Principal component regression (PCR) is an unpenalized regression
on principal components of all 94 characteristics, computed from the training
data alone; because it combines them linearly, the gap between PCR and the
neural models measures what nonlinearity adds.

\NNThree{} is GKX's three-hidden-layer network, with 32, 16 and 8 hidden units~\cite{gu2020}, batch
normalization and an \(l_1\) penalty, applied here to 94 inputs. It attains the
highest pooled out-of-sample \(R^2\) among the specifications they evaluate, and
deeper networks do not improve on it, so it is a demanding baseline. It is also
the closest comparison for the architecture claim: it sees the same
characteristics with the same flexibility, but its depth and widths are set by
hand. We do not include tree ensembles, since the question here is how a neural
architecture is specified rather than which family of models predicts best.

All neural models are trained with Adam~\cite{kingma2015} on mean squared error,
with batches of 10{,}000, at most 100 epochs and gradient clipping at one.
Training stops after ten epochs without validation improvement for the HNNs and
five for the dense networks, the HNNs being given longer because the sparse mask
slows convergence. To mitigate the variation induced by random initialization, each forecast averages ten runs with different seeds, following GKX~\cite{gu2020}. This matters here because the differences between architectures are small relative to that variation.

Hyperparameters are chosen by validation mean squared error every five test
years and held for the block, while graphs and weights are re-estimated
annually. Table~\ref{tab:grids} lists the grids. HNNs use no \(l_1\)
penalty, since their sparsity comes from the graph rather than from shrinkage.
Model selection is not eliminated but changes in kind: a search over depth and
width sequences becomes a search over one structural parameter and the learning
rate.

\begin{table}[h]
\centering
\caption{Hyperparameter grids. \(\eta\) is the learning rate,
\(\lambda_1\) the \(l_1\) penalty on dense weights, \(K\) the maximum clique
size, and \(\epsilon\) the Huber robustness parameter. The learning-rate grid is
\(\eta\in\{10^{-4},10^{-3},3\times10^{-3},10^{-2},3\times10^{-2}\}\).}
\label{tab:grids}
\small
\begin{tabular}{ll}
\toprule
Model & Grid\\
\midrule
\HNNMarg{}, \HNNMS{} & \(\eta\); \(K\in\{2,\ldots,7\}\)\\
\HNNShuffle{}        & \(\eta\)\\
\NNThree{}           & \(\eta\); \(\lambda_1\in\{10^{-6},10^{-5},10^{-3}\}\)\\
\MLPHNN{}            & \(\eta\); \(\lambda_1\in\{10^{-6},10^{-5},10^{-3}\}\)\\
Huber-3              & \(\epsilon\in\{1.1,1.35,1.5,2.0\}\)\\
PCR                  & components \(\in\{5,10,20,30,40,60,80,94\}\)\\
\bottomrule
\end{tabular}
\end{table}

\subsection{Evaluation and Inference}
\begin{table*}[!b]
\centering
\caption{Out-of-sample excess-return results, 1987--2016. \(R^2\) and monthly
gross spreads are percentages. ICs are means of monthly cross-sectional
correlations. Parameters are median trainable counts per ensemble member across
annual refits. Boldface marks the strongest estimate in each performance
column.}
\label{tab:main}
\small
\setlength{\tabcolsep}{3.5pt}
\begin{tabular}{lrrrrrrr}
\toprule
Model & $R^2_{\rm oos}$ & MAE & Pearson IC & Spearman IC
& EW spread & AW spread & Params.\\
\midrule
\input{tables/main_results_rows}
\end{tabular}
\end{table*}
Predictive accuracy is measured against a
forecast of zero,
\begin{equation}
 R^2_{\mathrm{oos}}=1-
 \frac{\sum_{i,t}(y^e_{i,t+1}-\widehat y^e_{i,t+1})^2}
      {\sum_{i,t}(y^e_{i,t+1})^2},
 \label{eq:r2}
\end{equation}
where the sums run over all observations in the test years. The benchmark is
zero rather than the historical mean because the mean return of an individual
stock is so noisy that it lowers the bar for good forecasting
performance~\cite{gu2020}.

We also report pooled mean absolute error, and the Pearson and Spearman
information coefficients: for each test month we compute the cross-sectional
correlation between predicted and realized returns, then average across months.
These matter because a long-short portfolio earns its return from ranking stocks
correctly within a month rather than from predicting the level of returns, and a
forecast that ranks better supports a higher risk-adjusted
return~\cite{grinold2000}.

Each month we sort stocks into deciles on the prediction and record the return
on the top decile minus the bottom. We report this under equal weights (EW) and
under lagged market equity (AW), since the two can differ substantially when
small stocks drive the result~\cite{famafrench2008}. Both legs earn the same
risk-free rate, so the raw and excess long-short spreads are identical.

Turnover is measured with a uniform transaction cost, since the measured
advantage of machine-learning strategies can shrink once trading frictions are
included~\cite{avramov2023}. Let \(r^{\mathrm{gross}}_t\) be the month's
long-short spread, \(w_{i,t}\) the signed target weight of stock \(i\), and
\(\widetilde w_{i,t-1}\) its weight after the previous month's returns but
before rebalancing. With each leg normalized to unit notional,
\begin{equation}
 \mathrm{TO}_t=\frac{1}{2}\sum_i
 |w_{i,t}-\widetilde w_{i,t-1}|,\qquad
 r^{\mathrm{net}}_t(c)=r^{\mathrm{gross}}_t-2c\,\mathrm{TO}_t ,
 \label{eq:cost}
\end{equation}
where \(c\) is the cost per dollar traded. The sum is the notional bought and
sold, so \(\mathrm{TO}_t\) is one-way turnover and the cost is \(c\) times twice
that. Dropping the first test month leaves 359 rebalances, and we report
\(c\in\{0,10,25,50\}\) basis points. The calculation ignores security-specific
spreads, short-borrow fees, market impact and capacity.

Models are compared month by month. For each test month we average the
difference in squared forecast errors across stocks, giving a monthly series
whose mean is tested with a Diebold--Mariano statistic~\cite{diebold1995} and
Newey--West heteroskedasticity- and autocorrelation-consistent (HAC) standard errors with six lags~\cite{neweywest1987}. The same paired
test is applied to absolute error, the information coefficients and the
portfolio returns; a negative loss difference favors the HNN, as does a positive
difference in coefficient or spread. Six measures, squared error, absolute
error, the two information coefficients and the two decile spreads, are compared
against three benchmarks (\NNThree{}, \HNNShuffle{}, \MLPHNN{}), and the
resulting 18 \(p\)-values are adjusted by Holm's procedure~\cite{holm1979}. We
call a difference significant at 5\% after adjustment.

\section{Results}\label{re}

\subsection{Predictive Performance}

Table~\ref{tab:main} reports the full comparison. \HNNMarg{} and \HNNMS{} match
\NNThree{} on pooled \(R^2_{\mathrm{oos}}\), at \HNNRtwo\% and \MSRtwo\% against
\NNRtwo\%. The \HNNMarg{} difference is not significant
(\(p=\HNNNNP\)), but the models separate more clearly on ranking. \HNNMarg{} has the highest
Pearson IC and the widest equal-weighted spread of any model, exceeding
\NNThree{}'s Pearson IC by 0.0056 (\(p=\HNNNNPearsonP\)), and this survives Holm adjustment (adjusted \(p=\HNNNNPearsonHolmP\)). The Spearman
difference points the same way without reaching significance (\(p=0.178\));
\NNThree{}'s MAE is marginally lower (\(p=0.050\)). Squared error compares each prediction with the realized return, and by that
standard the two models perform similarly. A decile sort depends only on the cross-sectional ordering of predicted returns, not on their absolute levels: it takes a long position in the top predicted decile and a short position in the bottom. The HNN's ranking advantage produces the stronger equal-weighted spread in
Table~\ref{tab:main}.

The three benchmarks differ in how many characteristics they use and whether
they combine them linearly. Going from Huber-3's three characteristics to PCR's
94 lifts pooled \(R^2\) from \(-0.137\)\% to 0.336\% and the equal-weighted
spread from 0.51\% to 2.87\%. Letting those 94 interact nonlinearly in
\NNThree{} raises them again to \NNRtwo\% and 3.75\%. Most of the predictive
power therefore comes from the number of characteristics and the freedom to
combine them, and what the HNN adds on top of \NNThree{} is ranking rather than
level accuracy. Huber-3 shows why both coefficients are reported. It has the highest Spearman IC
of any model but the lowest Pearson IC and the lowest equal-weighted spread.
Spearman IC evaluates rank agreement across the full cross-section, whereas the
decile spread depends specifically on the realized-return separation between
stocks assigned to the two extreme predicted deciles. Strong overall rank
correlation therefore need not imply a large long-short spread.

\begin{table*}[t]
\centering
\caption{Turnover and transaction-cost sensitivity over 359 monthly rebalances.
TO is mean total one-way turnover across two unit-notional legs. Net and SR are
the monthly net spread (percent) and annualized Sharpe ratio at 25 basis points
per traded dollar. BE is the break-even cost in basis points that sets the mean
spread to zero. EW and AW denote equal and lagged-market-equity weights.}
\label{tab:portfolio}
\normalsize
\setlength{\tabcolsep}{6pt}
\begin{tabular}{lrrrrrrrr}
\toprule
& \multicolumn{4}{c}{Equal weighted (EW)}
& \multicolumn{4}{c}{Asset weighted (AW)}\\
\cmidrule(lr){2-5}\cmidrule(lr){6-9}
Model & TO & Net & SR & BE & TO & Net & SR & BE\\
\midrule
\input{tables/economic_results_rows}
\end{tabular}
\end{table*}

\begin{figure*}[!b]
  \centering
  \includegraphics[width=\textwidth]{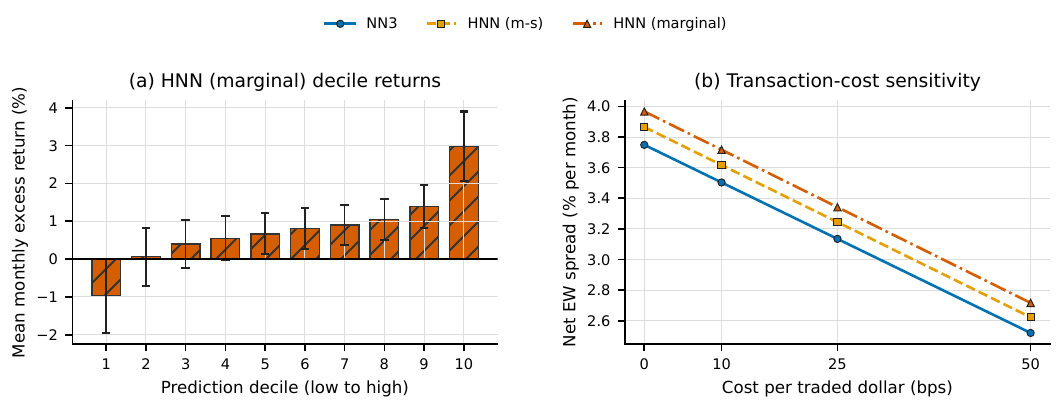}
  \caption{Economic ranking and cost sensitivity. Panel (a) reports \HNNMarg{}
  equal-weighted decile returns with 95\% HAC intervals (six lags). Panel (b)
  reports equal-weighted long-short spreads under uniform costs, using
  Equation~\eqref{eq:cost}.}
  \Description{The first panel shows HNN marginal realized returns increasing
  from low to high prediction deciles, with confidence intervals. The second
  shows three models' net long-short spreads declining as transaction costs
  increase; marker shapes and line styles distinguish the models.}
  \label{fig:economic}
\end{figure*}

\subsection{Feature Alignment and Parameter Economy}

The shuffled control changes which characteristic enters at which node and
nothing else: the widths, the connections and the parameter count are those of
\HNNMarg{}. Any difference is therefore due to the mismatch between the characteristics and their original positions in the MFCF. Pooled \(R^2\) falls from
\HNNRtwo\% to \ShuffleRtwo\%, the Pearson IC from 0.0642 to 0.0597, and the
equal-weighted spread from 3.95\% to 3.73\%. The Pearson reduction survives Holm
adjustment (adjusted \(p=\HNNShufflePearsonHolmP\)); the spread reduction is
significant before adjustment (\(p=\HNNShuffleEWP\)) but not after. What the filtered
graph contributes is therefore not only sparsity but also the particular clique structure it estimates.

The dense control instead replaces the HNN with a conventional width-matched dense architecture. \MLPHNN{} has the same widths as
\HNNMarg{} but connects every layer fully, which gives it \DenseParams{} million
parameters against \HNNMarg{}'s \HNNParams{} thousand. Taken as a complete
architecture, \HNNMarg{} nonetheless ranks the cross-section better on both
information coefficients and on the equal-weighted spread, and the Pearson
advantage again survives adjustment (adjusted \(p=\HNNDensePearsonHolmP\)).
\MLPHNN{} does have a slightly lower MAE and a higher asset-weighted spread, so
the HNN's gain is in the equal-weighted rather than the asset-weighted
portfolio.

Parameter count depends on the selected \(K\) and on how many subsets the
filtered graph contains. \HNNMS{} runs with \(K\le4\) in 20 of the 30 annual
refits, against \(K\ge5\) in 25 refits for \HNNMarg{}, which is why its median
count is lower even though it reaches \(K=7\) at the end of the sample. No layer
width is chosen by hand.

\subsection{Economic Ranking}

\begin{figure*}[t]
  \centering
  \includegraphics[width=\textwidth]{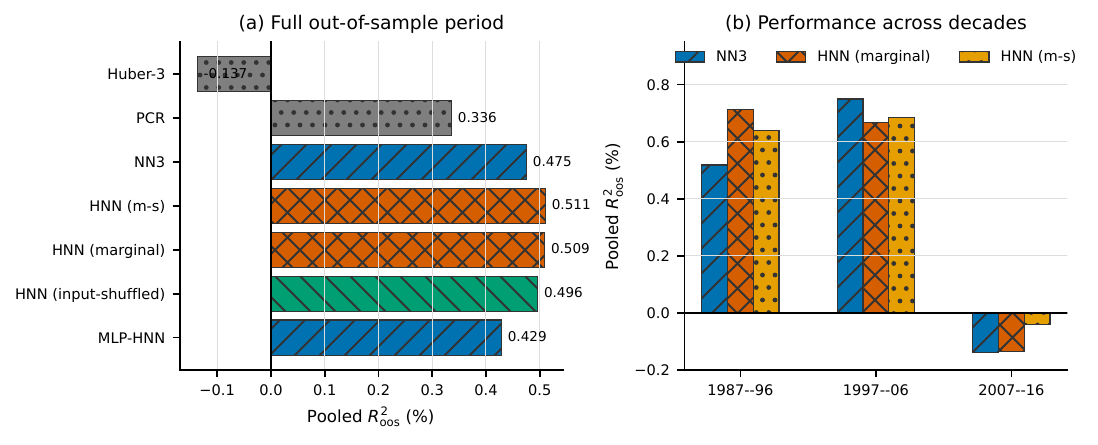}
  \caption{Pooled and decade-level predictive performance. Panel (a) reports
  full-period \(R^2_{\mathrm{oos}}\) for all seven models. Panel (b) reports the
  two HNN variants and \NNThree{} over three non-overlapping decades.}
  \Description{A horizontal bar chart compares full-period out-of-sample R
  squared across seven models. A grouped bar chart compares three selected
  models over three decades. Bar colors and hatch patterns distinguish model
  families.}
  \label{fig:performance}
\end{figure*}

Table~\ref{tab:portfolio} reports turnover and net performance. \HNNMarg{}'s
equal-weighted decile returns rise almost monotonically from \(-0.97\%\) in the
bottom decile to \(2.98\%\) in the top (Figure~\ref{fig:economic}), giving the
3.95\% gross spread. At a cost of 25 basis points per dollar traded the spread
is still \HNNNetEW\% per month, with an annualized Sharpe ratio of
\HNNNetEWSR{}, and the cost at which it would vanish is \HNNBreakEvenEW{} basis
points, about six times the level applied. 

The ranking of the models is unchanged across the cost levels. At 25 basis
points \HNNMarg{} earns \HNNNetEW\% against \NNNetEW\% for \NNThree{}, a
difference that is positive but not significant (\(p=\HNNNNNetP\)). It stays
ahead of \HNNShuffle{} and \MLPHNN{}, with unadjusted \(p=\HNNShuffleNetP\) and
\(p=\HNNDenseNetP\). The advantage over both controls therefore survives a
common turnover penalty.

Asset weighting reverses the comparison with \NNThree{}: \HNNMarg{} earns
\HNNNetAW\% against \NNNetAW\% (\(p=\HNNNNNetAWP\) in \NNThree{}'s favor).
Equal- and asset-weighted results diverge in this way when an effect is stronger
among smaller stocks~\cite{famafrench2008}.

\subsection{Performance Across Time}

No architecture leads in every decade (Figure~\ref{fig:performance}).
\HNNMarg{} records pooled \(R^2\) of 0.714\%, 0.668\% and \(-0.135\)\% over
1987--1996, 1997--2006 and 2007--2016, against \NNThree{}'s 0.518\%, 0.749\% and
\(-0.138\)\%, and leads \NNThree{} in \HNNWinYears{} of the 30 individual years
on \(R^2\) and in \HNNPearsonWinYears{} on Pearson IC. \HNNMS{} is the strongest
neural model in the last decade at \(-0.040\)\%, and PCR the only model with
positive \(R^2\) there, at 0.087\%.

Over 2007--2016 the equal-weighted spread is 2.55\% per month for \HNNMarg{} and
2.48\% for \NNThree{}. Level accuracy and cross-sectional ordering therefore come
apart in the later sample. The neural models lose to a forecast of zero while
their decile sorts still separate the cross-section.

\section{Discussion}\label{dis}

The two ablations narrow down what the filtered graph contributes. Permuting the characteristics across the nodes leaves the network
identical in size and shape, yet lowers the Pearson IC by an amount that survives multiplicity adjustment, so
the graph does more than impose sparsity, the particular cliques it forms carry
information. The width-matched dense comparator also produces a lower Pearson
IC despite having \ParamRatio{} times more parameters. Because this comparator
uses both fully connected layers and a conventional final-layer prediction head,
the comparison concerns the complete HNN architecture rather than sparse
connectivity in isolation. It nevertheless shows that replacing the HNN with a
substantially larger conventional dense architecture of the same induced depth
and widths does not improve cross-sectional ranking.

This has two practical consequences. First, tuning is limited to the clique
bound and the learning rate, instead of a search over depth and width. Second, each neural unit stands for a specific set of
characteristics, so the interactions a fitted model expresses can be read directly off the graph. A dense network of the same size offers no equivalent reading.

Some limits should be noted. The 94 characteristics exclude GKX's industry
indicators and macroeconomic interactions, so this is a controlled comparison of
architectures rather than a replication of their 920-input specification. The
net returns charge a single linear cost against measured turnover and ignore
security-specific spreads, short-borrow fees, market impact and capacity, each
of which can impact portfolio performance~\cite{avramov2023}.

\section{Conclusion}\label{con}

This paper uses estimated dependence among firm characteristics to specify a
forecasting architecture. The cliques returned by the MFCF become units in a
sparse neural network, and their inclusion relations determine the connections
between layers. Depth, width, and connectivity therefore follow from the
estimated dependence graph rather than being chosen separately. Across 30
annual out-of-sample tests on the GKX panel, the resulting HNNs forecast as accurately as a three-hidden-layer benchmark,
deliver stronger cross-sectional ranking performance, and use far fewer
parameters than a dense network with the same induced layer widths.

Two natural extensions remain. First, the dependence matrix could be estimated
using measures that capture nonlinear relationships or are more robust than
Pearson correlation. Second, \HNNMS{} could be generalized so that its
architecture varies with broader market regimes rather than remaining fixed
within each training window. More generally, the results show that empirical
dependence among firm characteristics can be used not only to summarize their
relationship structure, but also to specify a forecasting architecture.

\bibliographystyle{ACM-Reference-Format}
\bibliography{references}

\end{document}

%% file: tables/generated_numbers.tex
\newcommand{\NTest}{2,508,749}
\newcommand{\MSRtwo}{0.511}
\newcommand{\HNNRtwo}{0.509}
\newcommand{\NNRtwo}{0.475}
\newcommand{\ShuffleRtwo}{0.496}

\newcommand{\HNNNNP}{0.796}
\newcommand{\HNNNNPearsonP}{0.002}
\newcommand{\HNNNNPearsonHolmP}{0.034}

\newcommand{\HNNShufflePearsonHolmP}{0.003}
\newcommand{\HNNShuffleEWP}{0.007}

\newcommand{\HNNDensePearsonHolmP}{0.001}

\newcommand{\ParamRatio}{80}
\newcommand{\HNNParams}{21.0}
\newcommand{\DenseParams}{1.68}
\newcommand{\HNNWinYears}{19}
\newcommand{\HNNPearsonWinYears}{19}

\newcommand{\HNNNetEW}{3.34}
\newcommand{\HNNNetEWSR}{2.28}

\newcommand{\HNNBreakEvenEW}{159}
\newcommand{\HNNNetAW}{1.31}

\newcommand{\NNNetEW}{3.13}
\newcommand{\NNNetAW}{1.78}
\newcommand{\HNNNNNetP}{0.112}
\newcommand{\HNNNNNetAWP}{0.036}
\newcommand{\HNNShuffleNetP}{0.007}
\newcommand{\HNNDenseNetP}{0.028}

%% file: tables/main_results_rows.tex
Huber-3 & -0.137 & \textbf{0.1050} & 0.0227 & \textbf{0.0676} & 0.51 & 1.21 & 4\\
PCR & 0.336 & 0.1056 & 0.0437 & 0.0464 & 2.87 & 1.15 & 81\\
NN3 & 0.475 & 0.1057 & 0.0586 & 0.0474 & 3.75 & 2.44 & 3.8k\\
\HNNMS{} & \textbf{0.511} & 0.1058 & 0.0610 & 0.0491 & 3.85 & 2.00 & 5.2k\\
\HNNMarg{} & 0.509 & 0.1059 & \textbf{0.0642} & 0.0517 & \textbf{3.95} & 1.99 & 21.0k\\
\HNNShuffle & 0.496 & 0.1058 & 0.0597 & 0.0490 & 3.73 & 1.91 & 21.0k\\
MLP-HNN & 0.429 & 0.1057 & 0.0524 & 0.0436 & 3.47 & \textbf{2.48} & 1680.7k\\
\bottomrule

%% file: tables/economic_results_rows.tex
NN3 & 1.23 & 3.13 & 2.17 & 153 & 1.35 & 1.78 & 0.95 & 91\\
\HNNMS{} & 1.24 & 3.24 & 2.24 & 156 & 1.38 & 1.32 & 0.76 & 73\\
\HNNMarg{} & 1.25 & 3.34 & 2.28 & 159 & 1.36 & 1.31 & 0.73 & 73\\
\HNNShuffle & 1.26 & 3.12 & 2.11 & 149 & 1.41 & 1.22 & 0.65 & 68\\
MLP-HNN & 1.15 & 2.91 & 1.93 & 151 & 1.28 & 1.85 & 0.99 & 98\\
\bottomrule

%% file: references.bib
@article{gu2020,
  author  = {Gu, Shihao and Kelly, Bryan and Xiu, Dacheng},
  title   = {Empirical Asset Pricing via Machine Learning},
  journal = {The Review of Financial Studies},
  volume  = {33},
  number  = {5},
  pages   = {2223--2273},
  year    = {2020},
  doi     = {10.1093/rfs/hhaa009}
}

@article{freyberger2020,
  author  = {Freyberger, Joachim and Neuhierl, Andreas and Weber, Michael},
  title   = {Dissecting Characteristics Nonparametrically},
  journal = {The Review of Financial Studies},
  volume  = {33},
  number  = {5},
  pages   = {2326--2377},
  year    = {2020},
  doi     = {10.1093/rfs/hhz123}
}

@book{grinold2000,
  author    = {Grinold, Richard C. and Kahn, Ronald N.},
  title     = {Active Portfolio Management},
  edition   = {2},
  publisher = {McGraw-Hill},
  address   = {New York},
  year      = {2000}
}

@article{famafrench2008,
  author  = {Fama, Eugene F. and French, Kenneth R.},
  title   = {Dissecting Anomalies},
  journal = {The Journal of Finance},
  volume  = {63}, number = {4}, pages = {1653--1678}, year = {2008},
  doi     = {10.1111/j.1540-6261.2008.01371.x}
}

@book{kendall1970,
  author    = {Kendall, Maurice G.},
  title     = {Rank Correlation Methods},
  edition   = {4},
  publisher = {Griffin},
  address   = {London},
  year      = {1970}
}

@inproceedings{nair2010,
  author    = {Nair, Vinod and Hinton, Geoffrey E.},
  title     = {Rectified Linear Units Improve Restricted {B}oltzmann Machines},
  booktitle = {Proceedings of the 27th International Conference on Machine Learning},
  pages     = {807--814},
  address   = {Haifa, Israel},
  year      = {2010}
}

@book{cover2006elements,
  author    = {Cover, Thomas M. and Thomas, Joy A.},
  title     = {Elements of Information Theory},
  edition   = {2},
  publisher = {Wiley-Interscience},
  address   = {Hoboken, NJ},
  year      = {2006}
}

@article{tibshirani1996,
  author  = {Tibshirani, Robert},
  title   = {Regression Shrinkage and Selection via the Lasso},
  journal = {Journal of the Royal Statistical Society: Series B},
  volume  = {58},
  number  = {1},
  pages   = {267--288},
  year    = {1996}
}

@article{holm1979,
  author  = {Holm, Sture},
  title   = {A Simple Sequentially Rejective Multiple Test Procedure},
  journal = {Scandinavian Journal of Statistics},
  volume  = {6},
  number  = {2},
  pages   = {65--70},
  year    = {1979}
}

@article{huber1964,
  title={Robust Estimation of a Location Parameter},
  author={Huber, Peter J.},
  journal={The Annals of Mathematical Statistics},
  volume={35},
  number={1},
  pages={73--101},
  year={1964}
}

@article{chen2024,
  author  = {Chen, Luyang and Pelger, Markus and Zhu, Jason},
  title   = {Deep Learning in Asset Pricing},
  journal = {Management Science},
  volume  = {70},
  number  = {2},
  pages   = {714--750},
  year    = {2024},
  doi     = {10.1287/mnsc.2023.4695}
}

@article{kelly2019,
  author  = {Kelly, Bryan T. and Pruitt, Seth and Su, Yinan},
  title   = {Characteristics Are Covariances: A Unified Model of Risk and Return},
  journal = {Journal of Financial Economics},
  volume  = {134},
  number  = {3},
  pages   = {501--524},
  year    = {2019},
  doi     = {10.1016/j.jfineco.2019.05.001}
}

@article{kozak2020,
  author  = {Kozak, Serhiy and Nagel, Stefan and Santosh, Shrihari},
  title   = {Shrinking the Cross-Section},
  journal = {Journal of Financial Economics},
  volume  = {135},
  number  = {2},
  pages   = {271--292},
  year    = {2020},
  doi     = {10.1016/j.jfineco.2019.06.008}
}

@article{elsken2019,
  author  = {Elsken, Thomas and Metzen, Jan Hendrik and Hutter, Frank},
  title   = {Neural Architecture Search: A Survey},
  journal = {Journal of Machine Learning Research},
  volume  = {20},
  number  = {55},
  pages   = {1--21},
  year    = {2019},
  url     = {https://jmlr.org/papers/v20/18-598.html}
}

@article{tumminello2005,
  author  = {Tumminello, Michele and Aste, Tomaso and Di Matteo, Tiziana and Mantegna, Rosario N.},
  title   = {A Tool for Filtering Information in Complex Systems},
  journal = {Proceedings of the National Academy of Sciences},
  volume  = {102},
  number  = {30},
  pages   = {10421--10426},
  year    = {2005},
  doi     = {10.1073/pnas.0500298102}
}

@article{massara2017,
  author  = {Massara, Guido Previde and Di Matteo, Tiziana and Aste, Tomaso},
  title   = {Network Filtering for Big Data: Triangulated Maximally Filtered Graph},
  journal = {Journal of Complex Networks},
  volume  = {5},
  number  = {2},
  pages   = {161--178},
  year    = {2017},
  doi     = {10.1093/comnet/cnw015}
}

@misc{massara2019,
  author  = {Massara, Guido Previde and Aste, Tomaso},
  title   = {Learning Clique Forests},
  year    = {2019},
  howpublished = {arXiv:1905.02266},
  doi     = {10.48550/arXiv.1905.02266}
}

@inproceedings{wang2022icaif,
  author    = {Wang, Yuanrong and Aste, Tomaso},
  title     = {Network Filtering of Spatial-Temporal {GNN} for Multivariate Time-Series Prediction},
  booktitle = {Proceedings of the Third ACM International Conference on AI in Finance},
  pages     = {463--470},
  year      = {2022},
  publisher = {Association for Computing Machinery},
  address   = {New York, NY, USA},
  doi       = {10.1145/3533271.3561678}
}

@inproceedings{wang2023hnn,
  author    = {Wang, Yuanrong and Briola, Antonio and Aste, Tomaso},
  title     = {Homological Neural Networks: A Sparse Architecture for Multivariate Complexity},
  booktitle = {Proceedings of the Second Annual Workshop on Topology, Algebra, and Geometry in Machine Learning},
  series    = {Proceedings of Machine Learning Research},
  volume    = {221},
  pages     = {228--241},
  year      = {2023},
  publisher = {PMLR},
  address   = {Honolulu, HI, USA},
  url       = {https://proceedings.mlr.press/v221/wang23a.html}
}

@misc{lin2026,
  author  = {Lin, Hongyu and Briola, Antonio and Wang, Yuanrong and Aste, Tomaso},
  title   = {Compositional Sparsity as an Inductive Bias for Neural Architecture Design},
  year    = {2026},
  howpublished = {arXiv:2605.14764},
  doi     = {10.48550/arXiv.2605.14764}
}

@inproceedings{kingma2015,
  author    = {Kingma, Diederik P. and Ba, Jimmy},
  title     = {Adam: A Method for Stochastic Optimization},
  booktitle = {3rd International Conference on Learning Representations},
  year      = {2015},
  publisher = {OpenReview.net},
  address   = {San Diego, CA, USA},
  numpages  = {15},
  url       = {https://arxiv.org/abs/1412.6980}
}

@misc{ba2016,
  author  = {Ba, Jimmy Lei and Kiros, Jamie Ryan and Hinton, Geoffrey E.},
  title   = {Layer Normalization},
  year    = {2016},
  howpublished = {arXiv:1607.06450},
  doi     = {10.48550/arXiv.1607.06450}
}

@article{neweywest1987,
  author  = {Newey, Whitney K. and West, Kenneth D.},
  title   = {A Simple, Positive Semi-Definite, Heteroskedasticity and Autocorrelation Consistent Covariance Matrix},
  journal = {Econometrica},
  volume  = {55},
  number  = {3},
  pages   = {703--708},
  year    = {1987},
  doi     = {10.2307/1913610}
}

@article{diebold1995,
  author  = {Diebold, Francis X. and Mariano, Roberto S.},
  title   = {Comparing Predictive Accuracy},
  journal = {Journal of Business \& Economic Statistics},
  volume  = {13},
  number  = {3},
  pages   = {253--263},
  year    = {1995},
  doi     = {10.1080/07350015.1995.10524599}
}

@article{mantegna1999,
  author  = {Mantegna, Rosario N.},
  title   = {Hierarchical Structure in Financial Markets},
  journal = {The European Physical Journal B},
  volume  = {11},
  number  = {1},
  pages   = {193--197},
  year    = {1999},
  doi     = {10.1007/s100510050929}
}

@article{friedman2008,
  author  = {Friedman, Jerome and Hastie, Trevor and Tibshirani, Robert},
  title   = {Sparse Inverse Covariance Estimation with the Graphical Lasso},
  journal = {Biostatistics},
  volume  = {9},
  number  = {3},
  pages   = {432--441},
  year    = {2008},
  doi     = {10.1093/biostatistics/kxm045}
}

@article{poggio2017,
  author  = {Poggio, Tomaso and Mhaskar, Hrushikesh and Rosasco, Lorenzo and Miranda, Brando and Liao, Qianli},
  title   = {Why and When Can Deep---but Not Shallow---Networks Avoid the Curse of Dimensionality: A Review},
  journal = {International Journal of Automation and Computing},
  volume  = {14},
  number  = {5},
  pages   = {503--519},
  year    = {2017},
  doi     = {10.1007/s11633-017-1054-2}
}

@article{feng2020,
  author  = {Feng, Guanhao and Giglio, Stefano and Xiu, Dacheng},
  title   = {Taming the Factor Zoo: A Test of New Factors},
  journal = {The Journal of Finance},
  volume  = {75},
  number  = {3},
  pages   = {1327--1370},
  year    = {2020},
  doi     = {10.1111/jofi.12883}
}

@article{gu2021autoencoder,
  author  = {Gu, Shihao and Kelly, Bryan and Xiu, Dacheng},
  title   = {Autoencoder Asset Pricing Models},
  journal = {Journal of Econometrics},
  volume  = {222},
  number  = {1},
  pages   = {429--450},
  year    = {2021},
  doi     = {10.1016/j.jeconom.2020.07.009}
}

@article{harvey2016,
  author  = {Harvey, Campbell R. and Liu, Yan and Zhu, Heqing},
  title   = {\ldots{} and the Cross-Section of Expected Returns},
  journal = {The Review of Financial Studies},
  volume  = {29},
  number  = {1},
  pages   = {5--68},
  year    = {2016},
  doi     = {10.1093/rfs/hhv059}
}

@article{avramov2023,
  author  = {Avramov, Doron and Cheng, Si and Metzker, Lior},
  title   = {Machine Learning vs. Economic Restrictions: Evidence from Stock Return Predictability},
  journal = {Management Science},
  volume  = {69},
  number  = {5},
  pages   = {2587--2619},
  year    = {2023},
  doi     = {10.1287/mnsc.2022.4449}
}
